\documentclass[showpacs,nofootinbib,showkeys,eqsecnum,prd,aps,preprint]{revtex4-2}
\usepackage{amsfonts,amsmath,amssymb,bm,graphicx}

\begin{document}

\title{Photon emission due to vacuum instability under the action of a
quasi-constant electric field }
\date{\today }
\author{T. C. Adorno${}^{a}$}
\email{Tiago.Adorno@xjtlu.edu.cn}
\author{S.~P.~Gavrilov${}^{b,c}$}
\email{gavrilovsergeyp@yahoo.com}
\author{D.~M.~Gitman${}^{d,e}$}
\email{gitman@if.usp.br}

\begin{abstract}
Following a nonperturbative formulation of strong-field QED developed in our
earlier works, we consider photon emission accompanying vacuum instability
under the action of a quasi-constant strong electric field of finite
duration $T$. We construct closed formulas for the total probabilities and
study the photon emission accompanying an electron-positron pair creation
from a vacuum. We establish the domain of the applicability of the locally
constant field approximation (LCFA) for the photon emission. We study
angular and polarization distribution of the emission as well as emission
characteristics in a high-frequency approximations with respect of $1/T$.
The results presented in this work is suitable to a further development of
the LCFA proposed in [Phys. Rev. D 95, 076013 (2017)].
\end{abstract}


\affiliation{${}^{a}$Department of Physics, School of Mathematics and Physics, Xi'an
Jiaotong-Liverpool University, 111 Ren'ai Road, Suzhou Dushu Lake Science
and Education Innovation District, Suzhou Industrial Park, Suzhou 215123,
People's Republic of China;\\
${}^{b}$Department of Physics, Saint-Petersburg state forest technical university, 5  Institute lane, 194021, Saint-Petersburg, Russia;\\
${}^{c}$School of Physics and Engineering, ITMO University, 197101 St. Petersburg,
Russia;\\
${}^{d}$P.N. Lebedev Physical Institute, 53 Leninskiy prospect, Moscow,
Russia;\\
${}^{e}$Institute of Physics, University of S\~{a}o Paulo, SP. Brazil}

\maketitle

\section{Introduction}

In Quantum Electrodynamics (QED), the paradigm of constant and homogeneous
electromagnetic fields provides the theoretical foundations for the widely
used locally constant field approximation (LCFA); see, e.g. \cite%
{HebAG08,GiesK17,Karb17,APShab19,SAPShab21,GG17} and references therein.
Such an approximation is applicable to any quantum field theory (QFT) model
that involves\emph{\ }an interaction with the $U\left( 1\right) $ gauge
field ( \textquotedblleft electric-like field\textquotedblright ).
Consequently, strong-field QED methods can be systematically extended to
other models, which justifies the scientific significance of this approach
within a broader framework of quantum theories\emph{.} A strong constant
electric field ($\sim E_{c}=m^{2}c^{3}/e\hbar $, where $-e$\ is the electron
charge and $m$\ is the mass of an electron) can create electron-positron
pairs from a vacuum \cite{Schw51}. This phenomenon, known in the literature
as the Schwinger effect, has been extensively investigated in many
field-theoretic models, ranging from\emph{\ }high-energy physics,
astrophysics, and the\emph{\ }physics of nanostructures; see, for instance,
the recent review \cite{Adv-QED22}. In this article, we propose an
advancement of the LCFA method applicable to radiative processes accompanied
by\emph{\ }the Schwinger effect.

Pair creation from the vacuum induced by an external\emph{\ }electric field
is a transient phenomenon. Nevertheless, there exists a {window} in the
parameter range in which\emph{\ }the back-reaction of the created pairs on
the external field can be neglected (see Ref. \cite{GG08-a} for details).
Therefore, a physically meaningful formulation of the pair creation problem
presupposes an external field of finite time duration. Such a version of the
LCFA is proposed in Ref. \cite{GG17}. Using this approach, we consider the $%
3+1$ dimensional QED in the presence of the $T$-constant uniform electric
field that exists during a macroscopic large time period $T$, assumed to be
much larger than the characteristic time scale $\Delta t_{\mathrm{st}%
}=\left( eEc/\hbar \right) ^{-1/2}$, $E>0$. This field switches-on at $%
t_{1}=-T/2$ and -off at $t_{2}=T/2$ abrubtly and remains constant with
amplitude $E$ within this interval $\Delta t=t_{2}-t_{1}=T$. While
switching-on and -off effects of the external field may influence vacuum
instability in general \cite{AdoFerGavGit18}, these can be neglected if\ the
time interval $T$ is sufficiently large, namely%
\begin{equation}
T/\Delta t_{\mathrm{st}}>\max \left\{ 1,E_{c}/E\right\} \,.
\label{time-condition}
\end{equation}

Following a nonperturbative formulation of strong-field QED developed in
Refs. \cite{Git77,GavGit77,Gitman77,GavGitSh80,VGGS81} and summarized in the
book \cite{FGS}, we study the emission of a high-frequency photon ($\omega
\gg T^{-1}$) accompanying vacuum instability under the action of a
quasi-constant strong electric field $E$ of finite duration $T$. It is worth
noting that the probabilities amplitudes\ for the emission of a photon from
a single-electron (positron) state and electron-positron creation by a
photon in a constant electric field were previously considered by Nikishov
in Ref. \cite{Nikishov71} (see also Ref. \cite{Nikishov71b} where a photon
emission from the vacuum accompanied by an electron-positron pair creation
was considered). However,\emph{\ }these amplitudes present a satisfactory
description of photon emission only in the regime where the external field
is not very strong and the effect of the pair creation is tini. In the case
of a strong field, one has to use a perturbation theory with respect to the
radiative interaction for average values, which is quite different from the
technique that is\emph{\ }suitable for calculating amplitudes. In this
article we present an example of the perturbation theory with respect to the
radiative interaction for average values that allows incorporating the\emph{%
\ }LCFA.

The article is organized as follows: In Sect. \ref{S2} we present the
effective perturbation theory of the photon emission in the presence of a
strong electric field that is suitable for the LCFA. In Sect. \ref{S3} we
explicitly derive\ the total probability of one photon emission from the
vacuum in a constant electric field of finite duration. We establish the
domain of the applicability of the LCFA for the photon emission and show
that there are characteristic angular and polarization properties of the
emission in the range of high frequencies.

In what follows, we use the relativistic units $\hslash =c=1$\ in which the
fine structure constant is $\alpha =e^{2}/c\hslash =e^{2}$.

\section{Effective perturbation theory of the photon emission\label{S2}}

In the presence of the external fields that violate the vacuum stability,%
\emph{\ }radiative processes corresponding to the emission of a single
photon might occur either from an electron (a positron) or from the vacuum
accompanied by the creation of an electron-positron pair. In the framework
of the generalized Furry representation \cite{FGS}, the probability that a
photon with momentum $\mathbf{k}$ and polarization $\vartheta $ is emitted
from the vacuum accompanied by an electron-positron pair has the form:%
\begin{eqnarray}
&&\mathcal{P}_{1}\left( \mathbf{k}\vartheta |0\right) =\sum_{n^{\prime
},n}\left\vert w^{\left( 1\right) }\left( \overset{+}{n^{\prime }}\overset{-}%
{n};\mathbf{k}\vartheta |0\right) \right\vert ^{2}\left\vert
c_{v}\right\vert ^{2}\,,  \notag \\
&&w^{\left( 1\right) }\left( \overset{+}{n^{\prime }}\overset{-}{n};\mathbf{k%
}\vartheta |0\right) =\left\langle 0,\mathrm{out}|a_{n^{\prime }}\left(
\mathrm{out}\right) b_{n}\left( \mathrm{out}\right) c_{\mathbf{k}\vartheta
}S^{\left( 1\right) }|0,\mathrm{in}\right\rangle c_{v}^{-1}\,.  \label{et0}
\end{eqnarray}%
Here, the $a$'s, $b$'s, and $c$'s are annihilation operators of initial and
final electrons, positrons, and photons, respectively. Their adjoints are
creation operators. $c_{v}=\langle 0,\mathrm{out}|0,\mathrm{in}\rangle $\ is
the vacuum to vacuum transition amplitude. The indices $n=\left( \mathbf{p}%
,\sigma \right) $ and $n^{\prime }=\left( \mathbf{p}^{\prime },\sigma
^{\prime }\right) $ denote the complete set of particle's and antiparticle's
quantum numbers, where $\mathbf{p}$ is momentum and $\sigma =\pm 1$ is the
spin polarization. The external field under consideration\textrm{\ }is
directed along the $x$-axis and is described by the vector potential with
only one nonzero component:%
\begin{equation*}
A_{x}^{\mathrm{ext}}(t)=-E\left\{
\begin{array}{ll}
t_{1}, & t\in \mathrm{I}=(-\infty ,t_{1}),\ t_{1}=-T/2\, \\
t, & t\in \mathrm{Int}=[t_{1},t_{2}]\, \\
t_{2}, & t\in \mathrm{II}=(t_{2},\infty )\,,\ t_{2}=T/2\,%
\end{array}%
\right. .
\end{equation*}%
It is assumed that for $t<t_{1}$ and for $t>t_{2}$, the electric field is
absent, therefore the\emph{\ }initial $|0,\mathrm{in}\rangle $ and final $|0,%
\mathrm{out}\rangle $ are vacuum states of free $\mathrm{in}$- and $\mathrm{%
out}$- charged particles, respectively. These vacua are different due to a
difference of initial and final values of external electromagnetic field
potentials and $c_{v}=\left\langle 0,\mathrm{out}|0,\mathrm{in}\right\rangle
$ is the vacuum to vacuum transition amplitude. During the time interval $%
t_{2}-t_{1}=T$, the Dirac field interacts with the external field. Moreover,
in Eq. (\ref{et0}), $S^{\left( 1\right) }$ stands for the $S$-matrix
truncated at first-order with respect to the radiative interaction,%
\begin{eqnarray}
&&S=\mathcal{T}\exp \left[ -i\int \hat{\jmath}_{\mu }\left( x\right) \hat{A}%
^{\mu }\left( x\right) dx\right] \approx 1+S^{\left( 1\right) }\,,  \notag \\
&&S^{\left( 1\right) }=-i\int \hat{\jmath}_{\mu }\left( x\right) \hat{A}%
^{\mu }\left( x\right) dx\,,\ \ dx=dtd\mathbf{r}\,,\ \ d\mathbf{r}%
=dx^{1}dx^{2}dx^{3}\,,  \label{d2}
\end{eqnarray}%
wherein $\mathcal{T}$ denotes the time-ordering symbol, $\hat{\jmath}^{\mu
}\left( x\right) =-\left( e/2\right) \left[ \hat{\Psi}^{\dag }\left(
x\right) \gamma ^{0}\gamma ^{\mu },\hat{\Psi}\left( x\right) \right] $ is
the current density field operator, and $\gamma ^{\mu }$ are Dirac's
matrices. The Dirac field operators $\hat{\Psi}\left( x\right) $, $\hat{\Psi}%
^{\dag }\left( x\right) $, and the electromagnetic field operator $\hat{A}%
^{\mu }\left( x\right) $\ are in the interaction representation. The Dirac
field operators obey the Dirac equation with the potential $\mathbf{A}^{%
\mathrm{ext}}(t)$.

The $\mathrm{in}$- and $\mathrm{out}$- sets of creation and annihilation
operators\textrm{\ }of electrons and positrons\ are defined by the two
representations of the quantum Dirac field $\hat{\Psi}\left( x\right) $ as%
\begin{eqnarray}
\hat{\Psi}\left( x\right) &=&\sum_{n}\left[ a_{n}\left( \mathrm{in}\right) \
_{+}\psi _{n}\left( x\right) +b_{n}^{\dagger }\left( \mathrm{in}\right) \
_{-}\psi _{n}\left( x\right) \right] \,,  \notag \\
&=&\sum_{n}\left[ a_{n}\left( \mathrm{out}\right) \ ^{+}\psi _{n}\left(
x\right) +b_{n}^{\dagger }\left( \mathrm{out}\right) \ ^{-}\psi _{n}\left(
x\right) \right] \,,  \label{d4}
\end{eqnarray}%
where $\ _{\zeta }\psi _{n}\left( x\right) $ and $\ ^{\zeta }\psi _{n}\left(
x\right) $ are orthonormal $\mathrm{in}$- and $\mathrm{out}$-solutions of
the Dirac equation with the potential $\mathbf{A}^{\mathrm{ext}}(t)$. These
solutions have a well-defined sign of frequency $\zeta $ ($\zeta =+$ for
electrons and $\zeta =-$ for positrons) either before the field switches-on
or after it switches-off, respectively. They are\textrm{\ }related by a
linear transformation of the form%
\begin{eqnarray}
\ ^{\zeta }\psi _{n}\left( x\right) &=&g_{n}\left( _{+}|^{\zeta }\right)
\,_{+}\psi _{n}\left( x\right) +g_{n}\left( _{-}|^{\zeta }\right) \,_{-}\psi
_{n}\left( x\right) \,,  \notag \\
\ _{\zeta }\psi _{n}\left( x\right) &=&g_{n}\left( ^{+}|_{\zeta }\right)
\,^{+}\psi _{n}\left( x\right) +g_{n}\left( ^{-}|_{\zeta }\right) \,^{-}\psi
_{n}\left( x\right) \,,  \label{f7}
\end{eqnarray}%
where the decomposition coefficients are complex, $g_{n}\left( _{\zeta
^{\prime }}|^{\zeta }\right) =g_{n}\left( ^{\zeta }|_{\zeta ^{\prime
}}\right) ^{\ast }$. Because these coefficients obey certain unitarity
relations, all coefficients can be expressed in terms of two of them, e.g.
of $g\left( _{+}\left\vert ^{+}\right. \right) $ and $g\left( _{-}\left\vert
^{+}\right. \right) $. However, even the latter coefficients are not
completely independent as they satisfy the condition%
\begin{equation}
\left\vert g_{n}\left( _{-}\left\vert ^{+}\right. \right) \right\vert
^{2}+\left\vert g_{n}\left( _{+}\left\vert ^{+}\right. \right) \right\vert
^{2}=1\,.  \label{f9}
\end{equation}%
Then a linear canonical transformation (Bogolubov transformation) between $%
\mathrm{in}$- and $\mathrm{out}$- operators which follows from Eq. (\ref{d4}%
) is defined by these coefficients%
\begin{align}
a_{n}\left( \mathrm{out}\right) & =g_{n}\left( ^{+}|_{+}\right) a_{n}(%
\mathrm{in})+g_{n}\left( ^{+}|_{-}\right) b_{n}^{\dagger }(\mathrm{in}),
\notag \\
b_{n}^{\dagger }\left( \mathrm{out}\right) & =g_{n}\left( ^{-}|_{+}\right)
a_{n}(\mathrm{in})+g_{n}\left( ^{-}|_{-}\right) b_{n}^{\dagger }(\mathrm{in}%
).  \label{f10}
\end{align}

Using relations (\ref{f10}), one finds that the differential mean number of
the pairs created is
\begin{equation}
N_{n}^{\mathrm{cr}}=\left\vert g_{n}\left( {}_{-}|{}^{+}\right) \right\vert
^{2}.  \label{f14}
\end{equation}

The decomposition of the operator $\mathbf{\hat{A}}\left( x\right) $\ in
terms of creation and annihilation operators of free photons, $c_{\mathbf{k}%
\vartheta }^{\dag }$\ and $c_{\mathbf{k}\vartheta }$, reads:%
\begin{equation}
\mathbf{\hat{A}}(x)=\sum_{\mathbf{k,}\vartheta }\sqrt{\frac{2\pi }{V\omega }}%
\boldsymbol{\epsilon }_{\mathbf{k}\vartheta }\left[ c_{\mathbf{k}\vartheta
}\,e^{i(\mathbf{kr}-\omega t)}+c_{\mathbf{k}\vartheta }^{\dagger }\,e^{-i(%
\mathbf{kr}-\omega t)}\right] \,,  \label{d3}
\end{equation}%
where $\vartheta =1,2$\ denotes a polarization index, $\epsilon _{\mathbf{k}%
\vartheta }$\ are mutual orthogonal unit polarization vectors transversal to
a wave vector $\mathbf{k}$, $\omega =\left\vert \mathbf{k}\right\vert $, and
$V$ is the volume of the box regularization.

To construct a perturbation theory for the probability amplitude%
\begin{equation*}
W=\left\langle \mathrm{out}\left\vert \mathcal{S}\right\vert \mathrm{in}%
\right\rangle \,,
\end{equation*}%
for transition from an initial to a final state, one needs to reduce the $S$%
-matrix to a generalized normal form with respect to the vacua $\left\langle
0,\mathrm{out}\right\vert $ and $|0,\mathrm{in}\rangle $. This is exactly
how expression (\ref{et0}) was obtained. It presents a satisfactory
description of photon emission in the regime of weak external field, in
which the differential number of pairs created from the vacuum is naturally
small, $N_{n}^{\mathrm{cr}}=\left\langle 0,\mathrm{in}\left\vert
a_{n}^{\dagger }\left( \mathrm{out}\right) a_{n}\left( \mathrm{out}\right)
\right\vert 0,\mathrm{in}\right\rangle \ll 1$. Probabilities of such a kind
were considered by Nikishov for the case of a constant electric field \cite%
{Nikishov71b}. However, in the case of a strong field, emission processes
can be followed by the creation of\emph{\ }electron-positron pairs from the
vacuum. The resulting total probability of one photon emission from the
vacuum in this case admits a complicated representation. Nevertheless, this
difficulty can be circumvented by employing a perturbative approach
formulated for average values. Specifically, the generating functional of
mean values has the form%
\begin{equation*}
\left\langle \mathcal{F}\left( t\right) \right\rangle =\left\langle \mathrm{%
in}\left\vert \mathcal{S}^{-1}\mathcal{TF}\left( t\right) \mathcal{S}%
\right\vert \mathrm{in}\right\rangle \,,
\end{equation*}%
where the symbol $\mathcal{T}$ acts on both sides: it orders field operators
to the right of it and antiorders them to the left. For physically
admissible external fields, when the density of the number of pairs created
from a vacuum is finite and the final and initial vacuum are not orthogonal,
$\langle 0,\mathrm{out}|0,\mathrm{in}\rangle \neq 0$ and\ there is an
unitary operator $\mathcal{V}$ that relates the in and out- representations
of the Fock space via $|\mathrm{in}\rangle =V|\mathrm{out}\rangle $.

The total probability of one photon emission from the vacuum followed by any
number of created electron-positron pairs, $\mathcal{P}\left( \mathbf{k}%
\vartheta |0,\mathrm{in}\right) $, can be represented as a trace of the
operators $c_{\mathbf{k}\vartheta }\mathcal{S}\left\vert 0,\mathrm{in}%
\right\rangle \left\langle 0,\mathrm{in}\right\vert \mathcal{S}^{-1}c_{%
\mathbf{k}\vartheta }^{\dagger }$ with respect to the \textrm{out}-basis,
\begin{equation}
\mathcal{P}\left( \left. \mathbf{k}\vartheta \right\vert 0,\mathrm{in}%
\right) =\mathrm{tr\,}\left[ c_{\mathbf{k}\vartheta }\mathcal{S}\left\vert 0,%
\mathrm{in}\right\rangle \left\langle 0,\mathrm{in}\right\vert \mathcal{S}%
^{-1}c_{\mathbf{k}\vartheta }^{\dagger }\right] \ .  \label{f42b}
\end{equation}%
By using the unitary transformation $\mathcal{V}$, we can pass from the
basis of the final states to the basis of the initial states and represent
the trace (\ref{f42b}) as an average value of the photon number operator,%
\begin{equation}
\mathcal{P}\left( \left. \mathbf{k}\vartheta \right\vert 0,\mathrm{in}%
\right) =\left\langle 0,\mathrm{in}\right\vert \mathcal{S}^{-1}c_{\mathbf{k}%
\vartheta }^{\dagger }c_{\mathbf{k}\vartheta }\mathcal{S}\left\vert 0,%
\mathrm{in}\right\rangle \ ,  \label{ph_number}
\end{equation}%
where the $S$-matrix is truncated at first-order, $\mathcal{S}\approx
1+S^{\left( 1\right) }$. Generally speaking, one can find similar
expressions for emission probabilities when some charged particles are
already present in the initial state. For example, the total probabilities
of one photon emission from a single-electron state $\mathcal{P}\left(
\mathbf{k}\vartheta |\overset{+}{n}\right) $ can be presented as%
\begin{equation}
\mathcal{P}\left( \mathbf{k}\vartheta |\overset{+}{n}\right) =\left\langle 0,%
\mathrm{in}\left\vert a_{n}\left( \mathrm{in}\right) \mathcal{S}^{-1}c_{%
\mathbf{k}\vartheta }^{\dagger }c_{\mathbf{k}\vartheta }\mathcal{S}%
a_{n}^{\dagger }\left( \mathrm{in}\right) \right\vert 0,\mathrm{in}%
\right\rangle .  \label{ph_number-b}
\end{equation}

In course of constructing a perturbation theory with respect to the
radiative interaction for average values unlike the case of the probability
amplitudes\ one needs to reorganize the $S$-matrix in a normal form $:\ldots
:$ with respect to the \textrm{in}-vacuum. In the first-order approximation,
it is sufficient to represent only the operator $\mathbf{\hat{\jmath}}\left(
x\right) $ in such a form,%
\begin{equation}
\mathbf{\hat{\jmath}}\left( x\right) =\ :\mathbf{\hat{\jmath}}\left(
x\right) :+\ \langle \mathbf{j}\left( x\right) \rangle _{\mathrm{in}}\ ,\ \
\langle \mathbf{j}\left( x\right) \rangle _{\mathrm{in}}\ =\langle 0,\mathrm{%
in}\left\vert \mathbf{\hat{\jmath}}\left( x\right) \right\vert 0,\mathrm{in}%
\rangle \ .  \label{in-currentA}
\end{equation}%
The vacuum mean current $\langle \mathbf{j}\left( x\right) \rangle _{\mathrm{%
in}}$ is a sum of a vacuum polarization current and of a current of created
particles. It is a nontrivial object in a slowly varying electric field and
depends on the definition of the initial vacuum, $|0,\mathrm{in}\rangle $
and on the evolution of the electric field from the initial time\emph{\ }$%
t_{1}$\emph{\ }of switching on\emph{\ }to the time instant $t$\emph{.} After
switching off the electric field at time\emph{\ }$t_{2}$, the term $\langle
\mathbf{j}\left( x\right) \rangle _{\mathrm{in}}$ represents the current
density of the created pairs of charged particles. This current is a source
in the Maxwell equations for a mean electromagnetic field. Such a mean field
is a slowly varying crossed field emitted perpendicular to the axis of the
external electric field. In the frequency range of the photon emission $%
\omega \gg T^{-1},$ which is interesting to us, the contribution due to the
current $\langle \mathbf{j}\left( x\right) \rangle _{\mathrm{in}}$ can be
neglected. In the Fock space, the identity operator can be represented as a
sum of projection operators onto states with a certain number of initial
particles and antiparticles. Inserting this operator between the operators $%
c_{\mathbf{k}\vartheta }^{\dagger }$ and $c_{\mathbf{k}\vartheta }$, we can
represent the total probability (\ref{ph_number}) as follows:%
\begin{eqnarray}
&&\mathcal{P}\left( \mathbf{k}\vartheta |0\right) =\sum_{n^{\prime
},n}\left\vert w_{\mathrm{in}}^{\left( 1\right) }\left( \overset{-}{n}%
\overset{+}{n^{\prime }};\mathbf{k}\vartheta |0\right) \right\vert ^{2}\,,
\notag \\
&&w_{\mathrm{in}}^{\left( 1\right) }\left( \overset{-}{n}\overset{+}{%
n^{\prime }};\mathbf{k}\vartheta |0\right) =i\sqrt{\frac{2\pi }{V\omega }}%
\int \mathbf{j}_{\mathrm{in}}\left( \overset{-}{n}\overset{+}{n^{\prime }}%
|0\right) \boldsymbol{\epsilon }_{\mathbf{k}\vartheta }e^{i\left( \omega t-%
\mathbf{kr}\right) }dx\,,  \notag \\
&&\mathbf{j}_{\mathrm{in}}\left( \overset{-}{n}\overset{+}{n^{\prime }}%
|0\right) =\left\langle 0,\mathrm{in}\left\vert b_{n}\left( \mathrm{in}%
\right) a_{n^{\prime }}\left( \mathrm{in}\right) :\mathbf{\hat{\jmath}}%
\left( x\right) :\right\vert 0,\mathrm{in}\right\rangle \,.  \label{ga2.14}
\end{eqnarray}
Similarly, we obtain the total probability of one photon emission from an
electron.

Because the probability (\ref{ga2.14}) describes the process of one photon
emission from the vacuum,\emph{\ }we expect it is proportional to the total
number density of pairs produced, $n^{\mathrm{cr}}$. Assuming that the
electron density in the initial state is small, we see that the emission
from the vacuum provides the main contribution to the emission process.
Therefore, the contribution of processes with initial particles is not
considered here.

\section{LCFA\label{S3}}

\subsection{Photon emission in a constant electric field of finite duration
\label{SS3.1}}

In which follows, it is convenient to separate the components of the
momentum directed along the field and orthogonal to it as $\mathbf{p}=\left(
p_{x},\mathbf{p}_{\bot }\right) $. The electric field acting during the time
$T$ creates a considerable number of pairs from vacuum only in a finite
range in the momentum space,%
\begin{equation}
D:\Delta t_{st}^{2}p_{\bot }^{2}<T/\Delta t_{st}-\tau ,\;\Delta
t_{st}\left\vert p_{x}\right\vert <\frac{1}{2}T/\Delta t_{st}-\tau ,
\label{e3}
\end{equation}%
where $\tau $ is an arbitrary number satisfying the condition (see Ref. \cite%
{GG96} for details)%
\begin{equation}
T/\Delta t_{st}\gg \tau \gg \,\max \left\{ 1,E_{c}/E\right\} .  \label{Tcond}
\end{equation}%
In this range, the differential mean number of the pairs created is
identical with that of the constant electric field,
\begin{equation}
N_{n}^{\mathrm{cr}}=e^{-\pi \lambda }\,,  \label{N_cr}
\end{equation}%
and solutions of the Dirac equation can be represented as:%
\begin{eqnarray}
&&_{\pm }\psi _{n}\left( x\right) =\exp \left( i\mathbf{pr}\right) \;_{\pm
}\psi _{n}\left( t\right) ,  \notag \\
&&_{\pm }\psi _{n}\left( t\right) =\sqrt{eE}\left[ \left( \pm 1+i\right)
\;_{\pm }\varphi _{n,\mp 1}(t)+B\left( \mathbf{p}\right) \;_{\pm }\varphi
_{n,\pm 1}(t)\right] v_{\pm 1,\sigma },  \notag \\
&&B\left( \mathbf{p}\right) =\left( eE\right) ^{-1/2}\left( \alpha
^{2}p_{y}+\alpha ^{3}p_{z}+\gamma ^{0}m\right) ,\;\boldsymbol{\alpha }%
=\gamma ^{0}\boldsymbol{\gamma }\mathbf{,}  \notag \\
&&_{+}\varphi _{n,\varkappa }(t)=CD_{\nu -(1+\varkappa )/2}\left[ -(1-i)\xi %
\right] ,\ \ _{-}\varphi _{n,\varkappa }(t)=CD_{-\nu -(1-\varkappa )/2}\left[
-(1+i)\xi \right] \,,  \notag \\
&&\xi =\frac{eEt-p_{x}}{\sqrt{eE}},\;\nu =i\lambda /2,\;\lambda =\frac{%
p_{\bot }^{2}+m^{2}}{eE},\;C=\frac{\exp (-\pi \lambda /8)}{\sqrt{2eEV}}\,,
\label{2.10}
\end{eqnarray}%
where $D$'s are the linearly independent Weber parabolic cylinder functions
(WPCFs) \cite{HTF2}, and $v_{\varkappa ,\sigma }$ is a set of constant
orthonormalized spinors,
\begin{eqnarray}
\gamma ^{0}\gamma ^{1}v_{\varkappa ,\sigma } &=&\varkappa v_{\varkappa
,\sigma },\;\varkappa =\pm 1;  \notag \\
\;i\gamma ^{2}\gamma ^{3}v_{\varkappa ,\sigma } &=&\sigma v_{\varkappa
,\sigma },\ \sigma =\pm 1,\ v_{\varkappa ,\sigma }^{\dag }v_{\varkappa
^{\prime },\sigma ^{\prime }}=\delta _{\varkappa \varkappa ^{\prime }}\delta
_{\sigma \sigma ^{\prime }}\ .  \label{s1}
\end{eqnarray}%
We call the inequality (\ref{e3}) as the range of the stabilization for a
creation process.

Expressing the volume element $dk$ in spherical coordinates, $dk=\omega
^{2}d\omega d\Omega $, where $\omega $\ is the frequency of the radiated
photon in a region enclosed by the solid angle $d\Omega $, one can write the
probability of one photon emission with a given polarization $\vartheta $
per unit frequency and solid angle\textbf{,} which is accompanied by pair
production from the vacuum, as%
\begin{eqnarray}
&&\frac{d\mathcal{P}\left( \mathbf{k}\vartheta |0\right) }{d\omega d\Omega }%
=\alpha \frac{\omega \Delta t_{st}^{2}}{\left( 2\pi \right) ^{2}}\sum_{%
\mathbf{p}}\sum_{\sigma ,\sigma ^{\prime }=\pm 1}\left. \left\vert
M_{n^{\prime }n}^{0}\right\vert ^{2}\right\vert _{\mathbf{p}^{\prime }=%
\mathbf{p}-\mathbf{k}}\;,  \notag \\
&&M_{n^{\prime }n}^{0}=-\frac{V}{\Delta t_{st}}\int_{t_{1}}^{t_{2}}\;_{+}%
\bar{\psi}_{n^{\prime }}(t)\mathbf{\gamma }\boldsymbol{\epsilon }_{\mathbf{k}%
\vartheta }\;_{-}\psi _{n}(t)e^{i\omega t}dt\ ,  \label{ga4}
\end{eqnarray}%
where representation (\ref{d4}) is used and integral over the space volume $%
V $ is fulfilled.

In the stabilization range and neglecting switching-on and -off effects, one
can see that $M_{n^{\prime }n}^{0}$ is a linear combination of the following
integrals%
\begin{equation}
Y_{j^{\prime }j}^{0}\left( t_{2},t_{1}\right) =\int_{u_{1}}^{u_{2}}D_{-\nu
^{\prime }-j^{\prime }}\left[ -(1+i)u_{-}\right] D_{-\nu -j}\left[
-(1+i)u_{+}\right] e^{iu_{0}u}du,  \label{ga6}
\end{equation}%
where
\begin{eqnarray}
&&u=\Delta t_{st}\left[ eEt-\frac{1}{2}\left( p_{x}+p_{x}^{\prime }\right) %
\right] ,\;u_{1,2}=\left. u\right\vert _{t=t_{1,2}}\ ,  \notag \\
&&u_{x}=\Delta t_{st}\left( p_{x}^{\prime }-p_{x}\right) ,\;u_{\pm }=u\pm
u_{x}/2,\quad u_{0}=\Delta t_{st}\omega \ ,  \notag \\
&&\nu ^{\prime }=i\lambda ^{\prime }/2,\;\lambda ^{\prime }=\left. \lambda
\right\vert _{\mathbf{p}^{\prime }=\mathbf{p}-\mathbf{k}}\ .  \label{e10}
\end{eqnarray}

In the case of the $T$-constant electric field, there is the natural range
of the very low frequency of emission, $\omega \lesssim \omega ^{\mathrm{IR}%
}=2\pi T^{-1}$. In this range, generally speaking, the radiation must be
treated in the mean field approximation. We assume that the low energy
emission in this range does not exhaust the external field. Note that the
frequencies of soft photons, whose nature is associated with the
impossibility of separating a charged particle from its radiation field, are
lower than $\omega ^{\mathrm{IR}}$. For our purposes, it is enough to
restrict the applicability of the perturbation theory with respect of the
photon emission by the condition $\omega >\omega ^{\mathrm{IR}}$, which is
convenient to represent as:%
\begin{equation}
u_{0}>u_{0}^{\mathrm{IR}},\;u_{0}^{\mathrm{IR}}=2\pi \Delta t_{st}T^{-1}.
\label{IR1}
\end{equation}%
{\large \ }This condition provides the domain of the applicability of the $T$%
-constant electric field model in the framework of the perturbation theory.

The probability of the photon emission given by Eq. (\ref{ga4}), can be
integrated over $\mathbf{k}$ only between such limits that leave the
integral probability much smaller than unity. Let us demonstrate that for
the integration over $\omega $ there is a natural cutoff from above. Let us
consider the high frequency case,%
\begin{equation}
u_{0}\gtrsim \tau _{\gamma },  \label{lim3}
\end{equation}%
where $\tau _{\gamma }$ is an arbitrary given number, $\tau _{\gamma }\gg 1$%
. In the case under consideration, it is reasonable to assume that, for
example, $\tau _{\gamma }\gtrsim 3$. The probability (\ref{ga4}) is the
linear combination of integrals (\ref{ga6}). There are intervals where main
contributions to the integrals are formed. We can find this intervals using
the saddle-point method.

Let us consider integral $Y_{j^{\prime }j}^{0}\left( t_{2},t_{1}\right) $.
Under condition (\ref{lim3}), the mentioned saddle-point is situated in the
range where absolute values of arguments of both WPCF's involved in the
integral are big,
\begin{equation}
\left\vert u_{\pm }\right\vert \gg \max \left\{ 1,\lambda \right\} .
\label{lim3b}
\end{equation}%
In this case, if $u_{\pm }<0$, one uses the following asymptotic expansion
[(8.4.(1)) from Ref. \cite{HTF2}]:

\begin{equation}
D_{p}\left( z\right) =e^{-z^{2}/4}z^{p}\left[ 1+O\left( \left\vert
z\right\vert ^{-2}\right) \right] \;\mathrm{if}\;\left\vert \arg
z\right\vert <\frac{3\pi }{4}.  \label{asy_exp}
\end{equation}%
If $u_{\pm }>0$, one uses a relation between WPCF's (see (8.2.(7)) in Ref.
\cite{HTF2}) and then applies Eq. (\ref{asy_exp}). Thus one finds that the
saddle-point is $u=u_{0}/2$. Since $u_{0}$ is positive, the saddle-point can
be situated only in the range $u_{\pm }>0$.

In this range, the longitudinal kinetic momenta of a created electron is
negative, $P_{x}\left( t\right) =p_{x}-eEt<0$, and the longitudinal kinetic
momenta of a created positron is positive, $P_{x}^{\prime \left( p\right)
}\left( t\right) =-\left( p_{x}^{\prime }-eEt\right) >0$, and the modulus of
these momenta are large. This means that such electrons and positrons can be
treated as final particles, they acquire final longitudinal kinetic momenta
at the moment of time $t_{2}$. The kinetic energies of these particles are
determined mainly by their longitudinal kinetic momenta $\left\vert
P_{x}\left( t\right) \right\vert $ and $\left\vert P_{x}^{\prime }\left(
t\right) \right\vert $. We see that the saddle-point equation represents a
conservation law of the kinetic energy,
\begin{equation}
\left\vert P_{x}\left( t\right) \right\vert +\left\vert P_{x}^{\prime
}\left( t\right) \right\vert =\omega ,  \label{en_cons}
\end{equation}%
where $p_{x}^{\prime }=p_{x}-k_{x}$. In the neighborhood of the
saddle-point, $t=$ $t_{c}$, the corresponding kernels have Gaussian forms
with maxima at the time instant $t_{c}$,%
\begin{equation}
t_{c}=\frac{1}{2}\left( \Delta t_{st}^{2}\omega +\frac{p_{x}+p_{x}^{\prime }%
}{eE}\right) ,  \label{tc}
\end{equation}%
and with the standard deviation
\begin{equation}
\Delta t_{sd}=\Delta t_{st}/\sqrt{2}.  \label{sd}
\end{equation}%
The time $t_{c}$ corresponds to the position of the center of the formation
interval $\Delta t$ for given $\omega $, $p_{x}$, and $p_{x}^{\prime }$ in
the range $u_{\pm }>0$. The width of the formation interval $\Delta t$ must
be large enough to accommodate the points $u_{+}$ and $u_{-}$. In addition,
the formation interval must overlap the interval $\Delta t_{sd}$, $\Delta
t_{sd}<\Delta t$. It is natural to assume that $\Delta t\sim \Delta t_{st}$.
This\emph{\ }implies the following condition:%
\begin{equation}
\left\vert u_{x}\right\vert <1\ .  \label{lim4}
\end{equation}%
Under condition (\ref{lim3}) $\left\vert u_{x}\right\vert \ll u_{0}$, it
follows from Eq. (\ref{tc}) that%
\begin{equation}
t_{c}\approx \frac{1}{2}\Delta t_{st}^{2}\omega +\frac{p_{x}}{eE}\ .
\label{lim6}
\end{equation}%
This means that for the photon emission of a given frequency\emph{\ }$\omega
$, the dependence of the effect on\emph{\ }$p_{x}$\emph{\ }comes down to
just shifting of the center of the formation interval.\emph{\ }On the other
hand, for a given momentum $p_{x}$, photon with higher frequency is formed
later.

Thus, in the case of high frequencies, the width of the formation interval
does not depend on the frequency $\omega $ and on the momenta of the
particles and is determined entirely by the electric field $E$. The\
variation\ of\ the\ external electric\ field\ acting\ on\ the\ particle\
within\ the\ formation\ length\ can\ be\ neglected,\ which allows us to use
the LCFA.{\large \ }The obtained results can be easily extended to the study
of the emission in any slowly varying field configuration assuming that the
electric field $E\left( t\right) >0$ is uniform and time-dependent. In this
case, the kinetic energies in the conservation law (\ref{en_cons}) and
expressions derived from it have to be given by general forms,%
\begin{equation}
\left\vert P_{x}\left( t\right) \right\vert =\left\vert p_{x}+eA_{x}\left(
t\right) \right\vert ,\;\left\vert P_{x}^{\prime }\left( t\right)
\right\vert =\left\vert p_{x}^{\prime }+eA_{x}\left( t\right) \right\vert ,
\label{gen_kin}
\end{equation}%
where $A_{x}\left( t\right) $ is a potential step of a slowly varying field.
If the electric field decreases quickly enough beyond the formation
interval, the upper limitation to the intensity of the constant electric
field (see Ref. \cite{GG08-a} for details) can be significantly weakened.

It follows from Eq. (\ref{lim6}) that for any given $p_{x}$ the high
frequency emission, $\omega /\omega _{sc}\gtrsim \tau _{\gamma }$, $\omega
_{sc}=\Delta t_{st}^{-1}$, starts when the longitudinal kinetic momentum $%
P_{x}\left( t\right) $ reaching its threshold value at $t_{c}\sim t_{0}$
according to condition (\ref{lim3}),%
\begin{equation}
\frac{2\left\vert P_{x}\left( t_{0}\right) \right\vert }{\omega _{sc}}%
\approx \tau _{\gamma }\ .  \label{lim7}
\end{equation}%
The minimal frequency for the region of high frequencies is:
\begin{equation}
\omega _{\min }\approx \tau _{\gamma }\omega _{sc}\ .  \label{e34}
\end{equation}

The smallest possible value of the moment $t_{0}$ , at which Eq. (\ref{e34})
holds true, is achieved at the smallest possible momentum value $p_{x}$ from
the finite range (\ref{e3}). Taking it into account, we find%
\begin{equation}
t_{0}-t_{1}\sim \left( \tau _{\gamma }/2+\tau \right) \Delta t_{st}.
\label{e36}
\end{equation}

The frequency $\omega $ grows linearly, $\frac{d\omega }{dt_{c}}=2eE$, from
the minimum value $\omega _{\min }$ as long as the electric field is acting
and reaches the maximum possible for a given $p_{x}$ frequency $\omega _{2}$
at the time instant $t_{c}\sim t_{2}$, when the electric field switches off.
Photon with such a frequency is emitted during the formation interval
preceding the moment $t_{2}$ of switching off the electric field. It follows
from Eq. (\ref{lim6}) that%
\begin{equation}
\omega _{2}\approx 2\left\vert P_{x}\left( t_{2}\right) \right\vert .
\label{e32}
\end{equation}%
Absolute maximum among all possible frequencies $\omega _{2}$ with different
momenta $p_{x}$ satisfying Eq. (\ref{e3}) is:%
\begin{equation}
\omega _{\mathrm{\max }}\approx 2\Delta t_{st}^{-2}\left[ t_{2}+\max \left(
-p_{x}/eE\right) \right] \approx 2\left( T/\Delta t_{st}\right) \omega
_{sc}\ .  \label{e33}
\end{equation}%
A frequency range between $\omega _{\mathrm{\max }}$ and $\omega _{\min }$
does exists if
\begin{equation}
\frac{\omega _{\mathrm{\max }}}{\omega _{\min }}\approx \frac{2T}{\Delta
t_{st}\tau _{\gamma }}>1,  \label{e33a}
\end{equation}%
which means that the field duration time $T$ satisfying Eq. (\ref%
{time-condition}) and the upper limitation to the intensity is sufficiently
large.

We see that in the range of high-frequencies, the domain of the
applicability of the LCFA for a photon emission accompanying pair creation
from a vacuum is $\omega _{\min }<\omega <\omega _{\mathrm{\max }}$. The
obtained results can be extended to a general slowly varying field
configuration assuming that the kinetic energies in the conservation law (%
\ref{en_cons}) are given by general forms (\ref{gen_kin}). In this case,%
\begin{equation}
\omega _{\min }/\sqrt{eE\left( t_{0}\right) }=2\left\vert P_{x}\left(
t_{0}\right) \right\vert /\sqrt{eE\left( t_{0}\right) }\approx \tau _{\gamma
},\;\omega _{2}\approx 2\left\vert p_{x}+eA_{x}\left( t_{2}\right)
\right\vert >\omega _{\min }.  \label{lim11}
\end{equation}%
We find that the integral of the probability of one photon emission, given
by Eq. (\ref{ga4}), over frequency $\omega $ is bounded above by the value
of $\omega _{\mathrm{\max }}$.

\subsection{Large duration limit\label{SS3.2}}

For the momenta $p_{x}$ and $p_{x}^{\prime }$ satisfying condition (\ref{e3}%
) and for finite $u_{0}\ll \min \left( \left\vert u_{1}\right\vert
,\left\vert u_{2}\right\vert \right) $, one can use limit $T\rightarrow
\infty $ in integrals (\ref{ga6}). We denote the corresponding limits as:%
\begin{equation}
Y_{j^{\prime }j}^{0}\left( \rho ,\varphi \right) =\left. Y_{j^{\prime
}j}^{0}\left( t_{2},t_{1}\right) \right\vert _{T\rightarrow \infty }\ .
\label{e16}
\end{equation}%
Here, assuming $u_{0}^{2}-u_{x}^{2}\neq 0,$ the hyperbolic coordinates $\rho
$ and $\varphi $,%
\begin{eqnarray}
&&u_{0}=\rho \cosh \varphi ,\;u_{x}=\rho \sinh \varphi \ ,  \notag \\
&&\rho =\sqrt{u_{0}^{2}-u_{x}^{2}},\ \tanh \varphi =\frac{u_{x}}{u_{0}}\ ,
\label{e17}
\end{eqnarray}%
are introduced. We have $p_{x}^{\prime }=p_{x}-k_{x}$. Therefore, in any
frequency range the ratio $\left\vert u_{x}\right\vert /u_{0}$\ is bounded
above,
\begin{equation}
\frac{\left\vert u_{x}\right\vert }{u_{0}}=\frac{\left\vert k_{x}\right\vert
}{\omega }\leq 1.  \label{lim5}
\end{equation}%
The $\varphi $ dependence of integrals (\ref{e16}) can be factorized as (see
Appendix B in Ref. \cite{GG23} for details)
\begin{equation}
Y_{j^{\prime }j}^{0}\left( \rho ,\varphi \right) =e^{-i\beta \varphi
}Y_{j^{\prime }j}^{0}\left( \rho ,0\right) ,\;i\beta =\left( \nu -\nu
^{\prime }+j-j^{\prime }\right) /2,  \label{a6b}
\end{equation}

WPCF's in the integral $Y_{j^{\prime }j}^{0}\left( \rho ,0\right) $ are
solutions of the same differential equation. Taking it into account and
performing integrations by parts one finds that the function $Y_{j^{\prime
}j}^{0}\left( \rho ,0\right) $ satisfies the differential equation for the
confluent hypergeometric functions. An explicit form of $Y_{j^{\prime
}j}^{0}\left( \rho ,0\right) $ can be fixed by boundary conditions at $\rho
\rightarrow 0$ corresponding to the original integral; see Appendix B in
Ref. \cite{GG23} for details. The following relations take place:%
\begin{eqnarray}
&&\mathcal{J}_{j^{\prime }j}^{0}\left( \rho \right) =Y_{j^{\prime
}j}^{0}\left( \rho ,0\right) =e^{i\pi \left( \nu +\nu ^{\prime }+j+j^{\prime
}\right) /2}I_{j^{\prime }j}(\rho )\,,  \notag \\
&&I_{j^{\prime },j}(\rho )=\sqrt{\pi }e^{-i\pi /4}e^{i\rho ^{2}/4}Z^{i\beta
}\Psi \left( \nu +j,1+2i\beta ;Z\right) ,\ \ Z=e^{-i\pi /2}\rho ^{2}/2\ ,
\label{e21a}
\end{eqnarray}%
where $\Psi $ \ is the confluent hypergeometric function (CHF) (we use
notation of Ref. \cite{HTF1}).

We find that $M_{n^{\prime }n}^{0}$ has the form:%
\begin{eqnarray}
M_{n^{\prime }n}^{0} &\approx &-\frac{\mu }{2}\left[ 2i\tilde{\chi}_{\sigma
^{\prime }\sigma }^{\vartheta \left( 0,0\right) }Y_{00}^{0}\left( \rho
,\varphi \right) +\left( 1-i\right) \tilde{\chi}_{\sigma ^{\prime }\sigma
}^{\vartheta \left( 0,1\right) }Y_{01}^{0}\left( \rho ,\varphi \right)
\right.  \notag \\
&+&\left. \left( -1+i\right) \tilde{\chi}_{\sigma ^{\prime }\sigma
}^{\vartheta \left( 1,0\right) }Y_{10}^{0}\left( \rho ,\varphi \right) +%
\tilde{\chi}_{\sigma ^{\prime }\sigma }^{\vartheta \left( 1,1\right)
}Y_{11}^{0}\left( \rho ,\varphi \right) \right] ,  \label{am24a}
\end{eqnarray}%
where
\begin{eqnarray}
&&\tilde{\chi}_{\sigma ^{\prime }\sigma }^{\vartheta \left( 0,0\right)
}=v_{+1,\sigma ^{\prime }}^{\dag }\mathbf{\alpha }\boldsymbol{\epsilon }_{%
\mathbf{k}\vartheta }v_{-1,\sigma },\;\tilde{\chi}_{\sigma ^{\prime }\sigma
}^{\vartheta \left( 1,1\right) }=v_{+1,\sigma ^{\prime }}^{\dag }B\left(
\mathbf{p}^{\prime }\right) \mathbf{\alpha }\boldsymbol{\epsilon }_{\mathbf{k%
}\vartheta }B\left( \mathbf{p}\right) v_{-1,\sigma },  \notag \\
&&\tilde{\chi}_{\sigma ^{\prime }\sigma }^{\vartheta \left( 0,1\right)
}=v_{+1,\sigma ^{\prime }}^{\dag }\mathbf{\alpha }\boldsymbol{\epsilon }_{%
\mathbf{k}\vartheta }B\left( \mathbf{p}\right) v_{-1,\sigma },\;\tilde{\chi}%
_{\sigma ^{\prime }\sigma }^{\vartheta \left( 1,0\right) }=v_{+1,\sigma
^{\prime }}^{\dag }B\left( \mathbf{p}^{\prime }\right) \mathbf{\alpha }%
\boldsymbol{\epsilon }_{\mathbf{k}\vartheta }v_{-1,\sigma },  \notag \\
&&\mu =e^{-\pi \left( \lambda +\lambda ^{\prime }\right) /8}\exp \left(
i\omega \frac{p_{x}+p_{x}^{\prime }}{2eE}\right) .  \label{am25a}
\end{eqnarray}

To describe the angular distribution, we define the orthonormal triple%
\begin{eqnarray}
\mathbf{k/}k &=&(\cos \phi ,\,\sin \theta \sin \phi ,\,\cos \theta \sin \phi
)\,,  \notag \\
\boldsymbol{\epsilon }_{\mathbf{k}1} &=&\mathbf{e}_{x}\times \mathbf{k/}%
\left\vert \mathbf{e}_{x}\times \mathbf{k}\right\vert ,\quad \boldsymbol{%
\epsilon }_{\mathbf{k}2}=\mathbf{k\times }\boldsymbol{\epsilon }_{\mathbf{k}%
1}/\left\vert \mathbf{k\times }\boldsymbol{\epsilon }_{\mathbf{k}%
1}\right\vert ,  \label{ga7}
\end{eqnarray}%
such that%
\begin{eqnarray}
\boldsymbol{\epsilon }_{\mathbf{k}1} &=&(0,\,-\cos \theta ,\sin \theta \,)\,,
\notag \\
\boldsymbol{\epsilon }_{\mathbf{k}2} &=&(\sin \phi ,\,-\sin \theta \cos \phi
,-\,\cos \theta \cos \phi )\,,  \label{ga8}
\end{eqnarray}%
where $0\leq \phi \leq \pi $, $-\pi \leq \theta \leq +\pi $.

The matrix elements\emph{\ }$M_{n^{\prime }n}^{0}$ given\emph{\ }by Eq. (\ref%
{am24a}) involve complicated combinations of $\gamma $-matrices\emph{\ }%
making the\emph{\ }angular and polarization distributions of the emitted
photon difficult to analyse. \emph{\ }However, in the domain of the
applicability of the LCFA to the high frequency range (\ref{lim3}), $\omega
_{\min }<\omega <\omega _{\mathrm{\max }}$, the\emph{\ }analysis becomes
simpler.

To this end, we start with the fact that under condition (\ref{lim3}) $%
\left\vert u_{x}\right\vert \ll u_{0}$, it follows that $\tanh \varphi \ll 1$
then $\varphi \approx \cos \phi =k_{x}/\omega $, $\left\vert \cos \phi
\right\vert \ll 1$. We see that a contribution depending on parameter $%
\varphi $ in the integral (\ref{a6b}) is small and $u_{0}\approx \rho $. The
radiation is directed mainly near the plane orthogonal to the axis $x$.

In the case of the high frequency, $\rho \gtrsim \tau _{\gamma }\gg 1$, \
the both parameters $\ c$ and $Z$ of the CHF $\Psi \left( a,c;Z\right) $
appearing in Eq. (\ref{e21a}) are large and $a$ and the ratio $\eta =Z/c$
are fixed and positive, $\eta >0$. Under these conditions, using appropriate
asymptotic approximation (see sec. 13.8(ii) in Ref. \cite{DLMF}), we find
that the integral $Y_{j^{\prime }j}^{0}\left( \rho ,0\right) $ can be
approximated as%
\begin{eqnarray}
&&Y_{j^{\prime }j}^{0}\left( \rho ,0\right) \approx Ge^{i\pi \left(
j+j^{\prime }\right) /2}Z^{-j^{\prime }}\left[ D_{-\left( \nu +j\right)
}\left( 0\right) +O\left( \rho ^{-1}\right) \right] ,  \notag \\
&&G=\sqrt{\pi }e^{i\Theta }e^{-3\pi \left( \lambda +\lambda ^{\prime
}\right) /8},\;\Theta =\left[ \rho ^{2}-\left( \lambda +\lambda ^{\prime
}\right) \ln \rho ^{2}/2-\pi \right] /4,  \notag \\
&&D_{-j-\nu }\left( 0\right) =2^{-\left( \nu +j\right) /2}\sqrt{\pi }\Gamma
\left( 1/2+\left( \nu +j\right) /2\right) ^{-1}.  \label{ga9.1}
\end{eqnarray}%
The leading contributions to the amplitude $M_{n^{\prime }n}^{0}$ arise from
the terms proportional to $Y_{00}^{0}$ and $Y_{01}^{0}$,%
\begin{equation}
Y_{0j}^{0}\left( \rho ,\varphi \right) =e^{-i\varphi j/2}e^{\left( \lambda
-\lambda ^{\prime }\right) \varphi /4}Y_{0j}^{0}\left( \rho ,0\right) .
\label{ga9.2}
\end{equation}

Using the Dirac's representation for the $\gamma $-matrices, one finds:
\begin{eqnarray}
\tilde{\chi}_{\sigma ^{\prime }\sigma }^{\vartheta \left( 0,0\right) }
&=&\left\{
\begin{array}{ll}
0\,, & \mathrm{if}\ \ \sigma ^{\prime }=\sigma \, \\
\left( -i\epsilon _{\mathbf{k}\vartheta }^{2}+\sigma \epsilon _{\mathbf{k}%
\vartheta }^{3}\right) , & \mathrm{if}\ \ \sigma ^{\prime }\neq \sigma \,%
\end{array}%
\right. ,  \notag \\
\tilde{\chi}_{\sigma ^{\prime }\sigma }^{\vartheta \left( 0,1\right) } &=&%
\frac{\epsilon _{\mathbf{k}\vartheta }^{1}}{\sqrt{eE}}\left\{
\begin{array}{ll}
-m, & \mathrm{if}\ \ \sigma ^{\prime }=\sigma \, \\
\left( -ip_{y}+\sigma p_{z}\right) & \mathrm{if}\ \ \sigma ^{\prime }\neq
\sigma \,%
\end{array}%
\right. .  \label{ga9.2b}
\end{eqnarray}%
Then terms mixing\emph{\ }these matrix elements contribute only if $\sigma
^{\prime }\neq \sigma $ and $\vartheta =2$. In the leading-order term
approximation with respect to $\rho $, we find:%
\begin{equation}
\sum_{\sigma ,\sigma ^{\prime }=\pm 1}\left. \left\vert M_{n^{\prime
}n}^{0}\right\vert ^{2}\right\vert _{\mathbf{p}^{\prime }=\mathbf{p}-\mathbf{%
k}}\approx 2\pi e^{-\pi \left( \lambda +\lambda ^{\prime }\right) }\times
\left\{
\begin{array}{c}
\cosh \frac{\pi \lambda }{4},\;\mathrm{if}\;\vartheta =1 \\
\sinh \frac{\pi \lambda }{4}\left( 1+\mathcal{M}\cos \phi \right) ,\;\mathrm{%
if}\;\vartheta =2%
\end{array}%
\right. ,  \label{ga9.3}
\end{equation}%
where%
\begin{eqnarray*}
&&\mathcal{M}=1+2\pi \mathrm{Re}\Omega \left( \sinh \frac{\pi \lambda }{2}%
\right) ^{1/2}\left( \sinh \frac{\pi \lambda }{4}\right) ^{-1}, \\
&&\Omega =\frac{p_{\theta }}{\sqrt{eE}\sqrt{\lambda }}e^{i\Theta
_{M}},\;\Theta _{M}=-\frac{\pi }{4}-\arg \left[ \Gamma \left( \frac{1}{2}-%
\frac{i\lambda }{4}\right) \Gamma \left( 1+\frac{i\lambda }{4}\right) \right]
,
\end{eqnarray*}%
and remain the only terms linear with respect to $\cos \phi $ . Here the
projection of the vector $\mathbf{p}_{\bot }$ onto the direction of the
vector \ $\mathbf{k}_{\bot }$ is denoted as $p_{\theta }$. The projection of
this vector onto the perpendicular direction is denoted as $p_{\bar{\theta}}$%
, so that $\mathbf{p}_{\bot }^{2}=p_{\theta }^{2}+p_{\bar{\theta}}^{2}$.

We see that the probability of the high-frequency radiation given by Eqs. (%
\ref{ga4}) and (\ref{ga9.3}) has two linear polarizations. The polarization $%
\vartheta =2$ is in the plane $kX$ spanned by the direction of propagation
and the external electric field. The polarization $\vartheta =1$ is in the
plane orthogonal to the plane $kX$. \ In the case of polarization $\vartheta
=2$, the probability has a small contribition proportional to $\cos \phi
=k_{x}/\omega $. The quantity given by Eq. (\ref{ga9.3}) depends on the
frequency of the photon due to the parameter $\lambda ^{\prime }$,
\begin{equation}
\lambda ^{\prime }=\lambda +\rho ^{2}-2\rho p_{\theta }/\left( eE\right)
^{-1/2}.  \label{ga10}
\end{equation}

Now we can estimate the probability of the high-frequency emission, given by
Eq. (\ref{ga4}). In the case of the $T$-constant field, the momentum range $%
D $ is finite, given by Eq. (\ref{e3}). The probability (\ref{ga4}) can be
presented by an integral over the range $D$. Taking into account that the
contributions of large transverse momentum is exponentially suppressed, it
is sufficient to choose the limits of integration over its components as $%
-\varepsilon _{0}<p_{\theta }/\left( eE\right) ^{-1/2}<\varepsilon _{0}$ and
$-\varepsilon _{0}<p_{\bar{\theta}}/\left( eE\right) ^{-1/2}<\varepsilon
_{0} $, where $1\ll \varepsilon _{0}\ll \rho $. In this case, the
approximation (\ref{ga9.3}) is valid. Neglecting exponentially suppressed
terms and a small contribition depending on $\cos \phi $, we finally obtain
that%
\begin{eqnarray}
&&\frac{d\mathcal{P}\left( \mathbf{k}\vartheta |0\right) }{d\omega d\Omega }%
\approx \frac{\alpha n^{\mathrm{cr}}V}{16\pi ^{2}\omega _{sc}}\mathcal{R}%
\left( \mathbf{k,}\vartheta \right) ,  \notag \\
&&\mathcal{R}\left( \mathbf{k,}\vartheta \right) =e^{-\pi \left[ \lambda
_{0}+\rho ^{2}\right] }\left( \delta _{\vartheta 1}\cosh \frac{\pi \lambda
_{0}}{4}+\delta _{\vartheta 2}\sinh \frac{\pi \lambda _{0}}{4}\right) ,
\label{ga13}
\end{eqnarray}%
where $n^{\mathrm{cr}}$ is the density of the pairs created from the vacuum,%
\begin{equation}
n^{\mathrm{cr}}=r^{\mathrm{cr}}T,\;r^{\mathrm{cr}}=\frac{\left( eE\right)
^{2}}{4\pi ^{3}}e^{-\pi \lambda _{0}},\;\lambda _{0}=\frac{m^{2}}{eE}.
\label{ga14}
\end{equation}

We see that the dependence of the probability of the high-frequency emission
on frequency is determined mainly by the value of $e^{-\pi \rho ^{2}}$. In
the case of not very strong electric field, we have $\cosh \frac{\pi \lambda
_{0}}{4}\approx \sinh \frac{\pi \lambda _{0}}{4}\approx \exp \frac{\pi
\lambda _{0}}{4}$ then the probability of the emission for both polarization
are the same. In the case of a strong field, $\pi \lambda _{0}\lesssim 1$,
the emission is polarized in the plane orthogonal to the plane $kX$. The
emission accompanying the pair creation from vacuum is distinguished by the
fact of the cylindrical symmetry and its characteristic polarization
properties.

The obtained exact result for the large duration limit, given by Eqs. (\ref%
{a6b}) - (\ref{ga8}), can be also used as a basis for LCFA in the range of
a\ not very high frequencies, $1/T\ll \omega <\omega _{\min }$. We will
present the LCFA representation for the probability of the high-frequency
emission that hold true for any slowly varying field of a constant direction
somewhere else. Note that presented results can be useful to consider such
kind of the emission for 3D Dirac semimetals where the energy gap plays the
role of a mass term. When this gap is small enough, the critical field is
available in laboratory conditions.

\section{Conclusion\label{S4}}

Following a nonperturbative formulation of strong-field QED we present the
effective perturbation theory of the photon emission in the presence of a
strong electric field that is suitable to the locally constant field
approximation (LCFA). We construct closed formulas for the total
probabilities. We explicitly derive\ the total probability of one photon
emission from the vacuum in a constant electric field of finite duration $T$%
. We establish the domain of the applicability of the LCFA for the photon
emission and show that in the case of high frequencies, the width of the
formation interval does not depend on the frequency and on the momenta of
the particles and is determined entirely by the electric field $E$. The\
variation\ of\ the\ external electric\ field\ acting\ on\ the\ particle\
within\ the\ formation\ length\ can\ be\ neglected,\ which allows us to use
the LCFA.{\large \ }The frequency of the emission grows linearly and bounded
above by the value of the work of the electric field $2eET$. The radiation
is directed mainly near the plane orthogonal to the axis of the electric
field. The emission accompanying the pair creation from vacuum is
distinguished by the fact of the cylindrical symmetry and its characteristic
polarization properties. In the case of a strong field, the emission is
polarized in the direction  orthogonal to the wave vector and the electric
field. The obtained results can be easily extended to the study of the
emission in any slowly varying field configuration assuming that the
electric field $E\left( t\right) >0$ is uniform and time-dependent.

\subparagraph{\protect\large Acknowledgement}

D. M. G. thanks FAPESP (Grant No. 21/10128-0) and CNPq for permanent
support. The work of T. C. A. is funded by XJTLU Research Development
Funding, award no. RDF-21-02-056.

\subparagraph{\protect\large Conflict of Interest}

The authors declare that there is no conflict of interest, either existing
or potential.

\end{document}